\documentclass[aps,prd,preprint,groupedaddress]{revtex4}

\usepackage{bm}

\begin{document}
\title{PHOTON EMISSION IN A CONSTANT MAGNETIC FIELD IN 2+1 DIMENSIONAL SPACE-TIME}
\author{Jo\~ao Thiago de Santana Amaral}
\email[]{thiago.amaral@ufrgs.br}
\affiliation{Instituto de F\'{\i}sica, Universidade Federal do Rio Grande do Sul, Caixa Postal 15051, 91501-970 - Porto Alegre, RS, Brazil}
\author{Stoian Ivanov Zlatev}
\email[]{zlatev@fisica.ufs.br}
\affiliation{Departamento de Matem\'{a}tica, Universidade Federal de Sergipe, 49100-000 - S\~{a}o Crist\'{o}v\~{a}o, SE, Brazil}

\begin{abstract}
We calculate by the proper-time method the amplitude of the two-photon emission by a charged fermi\-on in a constant magnetic field in (2+1)-dimensional space-time.
The relevant dynamics reduces to that of a supersymmetric quantum-mechanical system with one bosonic and one fermionic degrees of freedom.
\end{abstract}
\maketitle

\section{Introduction}
The first-quantized formulation of particle theory has been proposed by Feynman `as an alternative to the formulation of second quantization' \cite{feynman50}. Its efficiency in calculations of one-loop boson scattering amplitudes has been fully recognized after the formulation of the Bern-Kosower rules 
\cite{Bern-Kosower91, Bern-Kosower92}. Within the Bern-Kosower approach, particle scattering amplitudes are obtained as infinite tension limits of some string amplitudes. Strassler has derived a similar set of rules `from first-quantized field theory' \cite{Strassler92}.  
The superior organization of the amplitudes, resulting in compact expressions, is an attractive feature of the
first-quantized (world-line) approach. It is important, in particular, in calculations of multi-particle amplitudes in gauge theories where the major difficulties lie in the great number of Feynman diagrams and terms. The five-gluon one-loop amplitude \cite{bdk93} and the four-graviton one-loop amplitude in quantum gravity \cite{bds93} have been calculated by Bern-Kosower method. 
Experimentally driven calculations of amplitudes in nonabelian gauge theories do not rely entirely on the world-line formalism. They are based nowadays on a large set of ideas and techniques (for a review see Ref.~\cite{bddk05}). Important results have been obtained recently
\cite{csw04,gk04,ggk04,wz04,kosower05, bddk05}
on the tree and one-loop amplitudes in nonabelian gauge theories, and, in particular, on the tree amplitudes with external fermions and scalars \cite{gk04,ggk04,wz04}. 

On the other hand, the world-line methods have been extremely useful in strong field calculations where the background field configuration cannot be treated as a small correction to a `trivial' one. Such calculations, when performed by standard field-theoretical methods, are often difficult even if the number of Feynman diagrams involved is not large. World-line techniques have been used in calculations of one-loop effective actions \cite{ss93, chd95, rss97, gmca98, shovkovy98, gs99, asz04}, as well as in computations of amplitudes in a strong field \cite{AS96, shai96, ds00, schubert00}. Multi-loop generalizations  \cite{Schmidt-Schubert94,Schmidt-Schubert96,dss96,roland-sato96,sato-schmidt98} have been used for calculations of the effective actions in QED \cite{rss97, ks99} and in Yang-Mills theory \cite{sato-schmidt99, ssz00}.  
The powerful techniques, based on the Bern-Kosower and Strassler's rules,
have been developed for processes without  fermions in the initial and final states and are not directly  applicable to scattering amplitudes between states containing fermions. The development of world-line techniques for these amplitudes was slower  \cite{McKR93, KK99, kk01}, although various  
path-integral (world-line) representations for the spinning particle propagator are known 
\cite{Fradkin65, Fradkin66, Barbashov65, BF70, HT82, Borisov-Kulish82, Polyakov87, Fainberg-Marshakov88, Fainberg-Marshakov88a, Fainberg-Marshakov90, ADJ90, Fradkin-Gitman91, Gitman-Saa93, AFP94}.  

Two-photon Compton scattering cross-sections \cite{Herold79, BAM86} and the two-photon emission rate \cite{SL99} in a constant magnetic field in $(3+1)$-dimensional space-time have been calculated by summation over the intermediate states. To our opinion, the world-line methods could be more promising, in particular, for generalizations to multi-photon processes.   

We calculate the amplitude (in the tree approximation) for  two-photon emission by a charged relativistic particle in a constant uniform  magnetic field in $2+1$-dimensional space-time within the world-line (first-quantized) framework. Schwinger proper-time integral is used. In the present paper, we use the operator method, although a path-integral formulation is also possible. The work on such a formulation is in progress.

The paper is organized as follows. In Sect.~2 we present the form of the amplitude for the two-photon emission in the constant magnetic field in $2+1$ dimensional space-time, as it comes in the quantum field theory. In Sect.~3 the one-fermion space is constructed along the lines of the Schwinger method \cite{Schwinger51}. In Sect.~4 we show that the relevant dynamics is that of a supersymmetric system with one bosonic and one fermionic degrees of freedom. In Sect.~5 an expression is derived for the amplitude. In the Appendix  the eigenstates of the Dirac hamiltonian are obtained along the lines of the Johnson-Lippmann construction \cite{JL49} in $3+1$ dimensions. 

\section{The $S$-matrix element}

In a constant magnetic field the vacuum is stable and the
three-potential $\mathcal{A}$ can be taken to be time-independent. Then Dirac hamiltonian
\begin{equation}
\label{hamiltonian0}
\hat H=-\boldsymbol{\alpha}\cdot\left(i\nabla+e\boldsymbol{\mathcal{A}}\right)
+m\gamma^0+e\mathcal{A}^0
\end{equation}
is time-independent. 
Let us denote by $\phi_N^{(+)}(\mathbf{x})$ and $\phi_N^{(-)}(\mathbf{x})$ (our notation here is close to that of Ref.~\cite{GFS90})
the positive and negative-energy eigenfunctions of the Dirac hamiltonian (\ref{hamiltonian0}), 
\[
\hat H\phi_N^{(+)}(\mathbf{x})=E_N^{(+)}\phi_N^{(+)}(\mathbf{x}), \qquad
\hat H\phi_N^{(-)}(\mathbf{x})=E_N^{(-)}\phi_N^{(-)}(\mathbf{x}), 
\]
where $E_N^{(+)}>0$, $E_N^{(-)}<0$ and $N$ stays for all the quantum numbers specifying the stationary state, excepting the energy sign. We assume that the eigenfunctions are normalized, 
\[
(\phi_M^{(+)}, \,\phi_N^{(+)})=(\phi_M^{(-)}, \,\phi_N^{(-)})=\delta_{MN}, 
\qquad (\phi,\,\chi)=\int\phi^\dagger(\mathbf{x})\chi(\mathbf{x})
\,d^2 x,
\]
then a completeness relation holds,
\[
\sum_N\left[\phi_N^{(+)}(\mathbf{x})\phi_N^{(+)\dagger}(\mathbf{y})
+\phi_N^{(-)}(\mathbf{x})\phi_N^{(-)\dagger}(\mathbf{y})\right]=\delta^2(\mathbf{x}-\mathbf{y}).
\]
Consider, for definiteness, the transition from a positive-energy state $\phi_{N_i}^{(+)}(\mathbf{x})$ to the posi\-tive-energy state $\phi_{N_f}^{(+)}(\mathbf{x})$  with emission of two photons with momenta $\mathbf{k}_1$ and $\mathbf{k}_2$. 
According to quantum field theory \cite{BD65}, the $S$-matrix element is given by
\[
S_{fi}=\frac{ie^2}{4\pi|\mathbf{k}_1||\mathbf{k}_2|}\left[R( \mathbf{k}_1, \mathbf{k}_2)+
R( \mathbf{k}_2, \mathbf{k}_1)
\right],
\]
where
\begin{eqnarray}
\label{R}
&&
R( \mathbf{k}_1, \mathbf{k}_2)=
\int d^3 x\int d^3y
\; \bar\phi_{N_f}^{(+)}(\mathbf{x})\frac{e^{i
(E_{N_f}^{(+)}+|\mathbf{k}_2|)x^0}}{\sqrt{2\pi}}e^{-i\mathbf{k}_2\cdot\mathbf{x}}\nonumber\\
&&\times\boldsymbol{\varepsilon}(\mathbf{k}_2)\cdot\boldsymbol{\gamma}\,S^{c}(x,y\mid \mathcal{A})\,
\boldsymbol{\varepsilon}(\mathbf{k}_1)\cdot\boldsymbol{\gamma}\,
\frac{e^{i(|\mathbf{k}_1|-E_{N_i}^{(+)})y^0}}{\sqrt{2\pi}}
e^{-i\mathbf{k}_1\cdot\mathbf{y}}\; \phi_{N_i}^{(+)}(\mathbf{y}),
\end{eqnarray} 
$S^{c}(x, y\mid \mathcal{A})$ is the fermion propagator in the external field,  
\begin{equation}
\label{propagator1}
\left[\gamma^\mu\left( i\partial-e\mathcal{A}\right)_\mu-m+i\epsilon
\right]S^c(x,y\mid \mathcal{A})=-\delta^3(x-y), 
\end{equation}
and $\boldsymbol{\varepsilon}(\mathbf{k})$ is the polarization vector associated with $\mathbf{k}$, 
\[
\mathbf{k}\cdot\boldsymbol{\varepsilon}(\mathbf{k})=0, \quad  
\mathbf{k}\wedge\boldsymbol{\varepsilon}(\mathbf{k})
\equiv{k}^1{\varepsilon}^2(\mathbf{k})-{k}^2{\varepsilon}^1(\mathbf{k}) =|\mathbf{k}|.
\]
As is known, there exist two irreducible two-dimensional representations of the Clifford algebra with three generating elements. The representations can be labelled by $s=\pm 1$ and chosen in such a way that the corresponding generators obey
\begin{equation}
\label{s}
[\gamma^\mu,\,\gamma^\nu]_-=-2is\epsilon^{\mu\nu\lambda}\gamma_\lambda.
\end{equation}
We omit the representation label $s$ in the sequel.

\section{The one-fermion space}

We represent, following Schwinger \cite{Schwinger51}, the propagator as a matrix element of an operator $S$ acting in a Hilbert space containing the (off-shell) one-fermion states (and no photons).

Consider a Hilbert space $\mathcal{H}(X,\mathcal{P})$ and a set of self-adjoint operators $X^\mu$, $\mathcal{P}_\mu$ $(\mu=0,1,2)$ satisfying the commutation relations
\[
[\mathcal{P}_\mu,  X^\nu]_-=i\delta_\mu^\nu \qquad 
[X^\mu,  X^\nu]_-=0, \qquad 
[\mathcal{P}_\mu,  \mathcal{P}_\nu]_-=0.
\]
We assume that the associative algebra, generated by those operators, acts irreducibly in $\mathcal{H}(X,\mathcal{P})$ and, moreover, 
\begin{equation}
\label{momenta}
\langle x \mid \mathcal{P}_\mu\mid \psi\rangle=i\partial_\mu
\langle x \mid \psi\rangle, \qquad \mid\psi\rangle\in \mathcal{H}(X,\mathcal{P}).
\end{equation}
where $|x\rangle$ are normalized common eigenvectors of $X^\mu$,  
\begin{equation}
\label{Xbasis}
X^\mu|x\rangle=x^\mu|x\rangle, \quad
\langle x\mid y\rangle=\delta^3(x-y), \quad 
\int d^3x|x\rangle\langle x|= I.
\end{equation}
Let $\mathcal{H}(\Gamma)$ be a two-dimensional complex space and
$\mid \rho\rangle$ ($\rho=1,2$) be vectors forming an orthonormal basis in $\mathcal{H}(\Gamma)$, 
\begin{equation}
\label{Gbasis}
\langle \rho\mid\rho^\prime\rangle=\delta_{\rho\rho^\prime}, \qquad 
\sum_{\rho=1}^2 \mid\rho\,\rangle\langle\,\rho\mid=I.
\end{equation}
The operators $\Gamma^\mu$, defined by 
$\langle \rho\mid \Gamma^\mu \mid\rho^\prime\rangle=\gamma^\mu_{\rho\rho^\prime}$,  satisfy
\begin{equation}
\label{Gammas}
[\Gamma^\mu, \Gamma^\nu]_+ =2\eta^{\mu\nu},\qquad
[\Gamma^\mu, \Gamma^\nu]_- =-2is\epsilon^{\mu\nu\lambda}\Gamma_\lambda. 
\end{equation}
The state vector in the one-particle space $\mathcal{H}=\mathcal{H}(X,\mathcal{P})\otimes\mathcal{H}(\Gamma)$, corresponding to the wave function $(2\pi)^{-1/2}e^{-ip_0x^0}\phi_{N}^{(+)}(\mathbf{x})$ is given by 
\begin{equation}
\mid p_0; E>0, N\rangle=\frac 1 {\sqrt{2\pi}}\int d^3x
\sum_{\rho=1}^{2}\mid x, \rho\rangle \, \left[\phi_N^{(+)}\right]_\rho(\mathbf{x})e^{-ip_0 x^0}.
\end{equation}  
It is, generally, an off-shell state since the value of  $p^0$ is arbitrary.
The propagator is a matrix element of an operator $S$ in $\mathcal{H}$,
\[
S^c_{\rho\rho^\prime}(x,y)=\langle x, \rho\mid S \mid y, \rho^\prime\rangle, 
\] 
Using eqs.~(\ref{propagator1}), (\ref{momenta}), (\ref{Xbasis}), and (\ref{Gbasis}), one finds that $S$ satisfies  
\begin{equation}
\label{S1}
\left(m-\not\!\Pi\right)S=I,
\end{equation}
where $\not\! \Pi =\Gamma^\mu\Pi_\mu$ and $\Pi^\mu$ are the operators of the kinematic momenta, 
\begin{equation}
\label{kinematic}
\Pi^\mu = \mathcal{P}^\mu-e  \mathcal{A}^\mu.
\end{equation}
Using the corresponding completeness relations for the bases, one can put eq.~(\ref{R}) into the form: 
\begin{eqnarray}
\label{R1}
&&
R(\mathbf{k}_1, \mathbf{k}_2)\\
&&=
\langle p^0_f; E>0,\, N_f\mid\nonumber \Gamma^0e^{-i\mathbf{k}_2\cdot \mathbf{X}}
\boldsymbol{\varepsilon}(\mathbf{k}_2)\cdot\boldsymbol{\Gamma}\,S\,
\boldsymbol{\varepsilon}(\mathbf{k}_1)\cdot\boldsymbol{\Gamma}\, e^{
-i\mathbf{k}_1\cdot\mathbf{X}} \mid p^0_i; E>0,\, N_i\rangle, 
\nonumber
\end{eqnarray} 
where
\[
p^0_f=E_{N_f}^{(+)}+|\mathbf{k}_2|,\qquad   p^0_i=E_{N_i}^{(+)}-|\mathbf{k}_1|.
\]
\section{The superoscillator}

Using the Schwinger proper-time integral \cite{Schwinger51}, we  represent the operator $S$ (in a general external field $\mathcal{A}$) as 
\begin{equation}
\label{S2}
S=i(m+\not\!\Pi)\int_0^\infty d\lambda \exp\left[
-i\lambda(m^2-\not \!\Pi^2)\right].
\end{equation}
The factor $(m+\not\!\Pi)$ in eq.~(\ref{S2}) can be expressed as an integral over a Grassmann variable $\chi$,
\begin{equation}
\label{grassmann}
m+\not\!\Pi=i\int d\chi(1-i\chi m)e^{-i\chi\not\Pi}, 
\end{equation}
where we assume that $\chi$ anticommutes with the $\Gamma^\mu$ and commutes with the rest of the operators.
Substituting eq.~(\ref{grassmann}) in eq.~(\ref{S2}) one obtains
\begin{equation}
\label{S3}
S=\int d\chi (i\chi m-1)\int_0^\infty d\lambda\,\exp(-i\lambda m^2) \,\exp\left[-i\chi\not\!\Pi
+i\lambda\not \!\Pi^2\right].
\end{equation}
The operator 
\begin{equation}
\label{U}
U=\exp\left[-i\chi\not\!\Pi+i\lambda\not \!\Pi^2\right]
\end{equation}
can be considered as an evolution operator (on the ``time" interval $0\le\tau\le 1$) of a quantum-mechanical system
with  the hamiltonian $\chi\not\!\!\Pi-\lambda\not \!\!\Pi^2$, containing a Grassmann-odd parameter $\chi$. The system is rather simple if the external field is a constant magnetic one. We choose in what follows
\begin{equation}
\label{potential}
\mathcal{A}^0=0, \qquad
\mathcal{A}^1=-\frac{Bx^2}{2}, \qquad 
\mathcal{A}^2= \frac{Bx^1}{2}.
\end{equation}
The magnetic field strength $B$
is assumed positive.
The operators $\Pi_\mu$ satisfy the commutation relations
\begin{equation}
\label{crkinematic}
[\Pi^k, \,\mathcal{P}^0]_-=0,\qquad
[\Pi^1, \,\Pi^2]_-=ieB.
\end{equation}
We choose, for definiteness, $e>0$ and make a linear canonical change.
The  operators $X^0$, $-\mathcal{P}^0$,
\begin{equation}
\label{QP}
Q=\frac {\sqrt{2}} b \Pi^1, \qquad P=\frac {\sqrt{2}} b \Pi^2, 
\end{equation}
and
\begin{equation}
\label{Q2P2}
\tilde Q=\frac {\sqrt{2}} b \left(\mathcal{P}^2+\frac{b^2}{4}X^1\right), \qquad \tilde P=\frac {\sqrt{2}} b \left(\mathcal{P}^1-\frac{b^2}{4}X^2\right),  
\end{equation}
where $b=\sqrt{2e B}$, are canonical, i.e., the following commutation relations hold   
\begin{equation}
\label{CCR}
[\mathcal P^0,\, X^0]_-=i,\qquad [Q, \, P]_-=i, \qquad [\tilde Q, \, \tilde P]_-=i, 
\end{equation}
the rest of the commutators being equal to zero.
In the three-potential (\ref{potential}) the operator $U$, eq.~(\ref{U}), factorizes, 
\begin{eqnarray}
\label{U1}
&&U=U_0 U_1, \qquad
U_0=\exp\left\{-i\chi \Gamma^0\mathcal{P}_0+i\lambda\mathcal{P}_0^2
\right\}, \\
\label{U-1}
&&U_1=\exp\left\{i\frac{\chi b}{\sqrt{\,2}}\left(\Gamma^1Q+\Gamma^2 P\right)
+i\frac{\lambda b^2}{2}\left(\Gamma^1Q+\Gamma^2 P\right)^2\right\}.
\end{eqnarray}
The operator $U_1$ acts in the factor $\bar\mathcal{H}=\mathcal{H}(Q,P)\otimes\mathcal{H}(\Gamma)$ of the tensor-product space $\mathcal{H}$. It
can be considered as an evolution operator (on the ``time" interval $0\le\tau\le 1$) of a quantum-mechanical system with one bosonic and one fermionic degrees of freedom. The hamiltonian
\[
h=-\frac {\chi b}{\sqrt{\,2}}\left(\Gamma^1Q+\Gamma^2 P\right)-
\frac{\lambda b^2}{2}\left(\Gamma^1Q+\Gamma^2 P\right)^2,
\]
contains the Grassmann-odd parameter $\chi$.  
The operators
\begin{equation}
\label{creation}
a=\frac 1 {\sqrt{\,2}}(Q+iP) \qquad 
a^\dagger=\frac 1 {\sqrt{\,2}}(Q-iP) 
\end{equation}
are (bosonic) lowering and raising operators, 
\begin{equation}
\label{bosonic}
[a,\, a^\dagger]_-=1,
\end{equation}
while the operators 
\begin{equation}
\label{creationF}
\alpha=\frac 1 {2i}(\Gamma^1+i\Gamma^2), \qquad
\alpha^\dagger=\frac 1 {2i}(\Gamma^1-i\Gamma^2)
\end{equation}
are fermionic ones, 
\begin{equation}
\label{anti}
[\alpha,\, \alpha^\dagger]_+=1, \qquad \alpha^2=(\alpha^\dagger)^2=0.
\end{equation}
Using eq.~(\ref{Gammas}), one finds
\begin{equation}
\label{comm}
[\alpha^\dagger,\Gamma^0 ]_-=2s\alpha^\dagger, \qquad
[\alpha,\Gamma^0 ]_-=-2s\alpha, \qquad
[\alpha,\alpha^\dagger ]_-=s\Gamma^0.
\end{equation}
When expressed in terms of these operators, Dirac operator in the potential (\ref{potential}) reads
\begin{equation}
\not\!\Pi-m=\Gamma^0\mathcal{P}_0-ib\left(
\alpha a^\dagger+\alpha^\dagger a
\right)
\end{equation}
and  the hamiltonian $h$ in the proper-time representation is given by
\begin{equation}
\label{h2}
h=\lambda b^2(a^\dagger a+\alpha^\dagger\alpha)-i\chi b(\alpha a^\dagger+\alpha^\dagger a).
\end{equation}
The hamiltonian $h$, as well as $H$,  eq.~(\ref{hamQP}), are supersymmetric. The nilpotent operators $K=\alpha a^{\dagger}$, $K^{\dagger}$ and the operator $h_{SUSY}=a^\dagger a+\alpha^{\dagger}\alpha$ (the hamiltonian for the supersymmetric oscillator \cite{Nicolai76,BM75,BDZVH76,Witten81,SH82}) generate a Lie superalgebra. The nonvanishing (anti-)commutator is  
\[
[K, K^{\dagger}]_{+}=h_{SUSY}.
\]
The hamiltonian $h$ is invariant with respect to the supertransformations generated by $K_{-}=K-K^{\dagger}$ only,  
\[
[K_{-}, h]=0.
\]
The existence of a natural supersymmetric structure associated with the Dirac equation is known \cite{BM75,BCL76,BDZVH76, Witten81} for a long time. We note that the hamiltonian (\ref{h2}) possesses only a residual supersymmetry (with respect to transformations generated by $K_{-}$) if compared with the supersymmetric oscillator.

Finally, using eqs.~(\ref{S3}), (\ref{U1}) and (\ref{U-1}), one obtains the following expression for the operator $S$ in a constant magnetic field:
\begin{eqnarray}
\label{S4}
S&=&\int d\chi \int_0^\infty d\lambda\,(i\chi m-1)\exp\left[-i\lambda (m^2-\mathcal{P}_0^2)-i\chi\Gamma^0\mathcal{P}_0\right]\nonumber\\
&&\times\exp\left\{-i\left[
\lambda b^2(a^\dagger a+\alpha^\dagger\alpha)-i\chi b(\alpha a^\dagger+\alpha^\dagger a)
\right]\right\}.
\end{eqnarray}

\section{The two-photon emission amplitude}
Using the orthonormality of the basis $\{|p^0\rangle\}$ in $\mathcal{H}(X^0, \mathcal{P}^0)$, and taking into account that the states in eq.~(\ref{R1}) are on-shell ones, one can represent $R(\mathbf{k}_1, \mathbf{k}_2)$ as a product of the energy-conservation $\delta$-function and a matrix element in the space
$\mathcal{H}(Q, P)\otimes\mathcal{H}(\tilde Q, \tilde P)\otimes\mathcal{H}(\Gamma)$,
\begin{eqnarray}
R(\mathbf{k}_1, \mathbf{k}_2)=\delta(E_{N_f}^{(+)})+|\mathbf{k}_1|+
|\mathbf{k}_2|-E_{N_i}^{(+)})T^{(2,1)}_{fi}, 
\end{eqnarray}
where 
\begin{eqnarray*}
T^{(2,1)}_{fi}
=\langle  E>0,\, N_f\mid
e^{-i\mathbf{k}_2\cdot\mathbf{X}} W(\mathcal{E}) 
e^{-i\mathbf{k}_1\cdot\mathbf{X}}
\mid  E>0,\, N_i\rangle,
\end{eqnarray*}
and
\begin{eqnarray*}
W(\mathcal{E})&=&\Gamma^0\boldsymbol{\epsilon}(\mathbf{k}_2)\cdot\boldsymbol{\Gamma}
\left[m-\Gamma^0\mathcal{E}+\frac{b}{\sqrt{\,2}}\left(
\Gamma^1 Q+\Gamma^2 P
\right)\right]^{-1}
\boldsymbol{\epsilon}(\mathbf{k}_1)\cdot\boldsymbol{\Gamma}\\
&=&\Gamma^0\boldsymbol{\epsilon}(\mathbf{k}_2)\cdot\boldsymbol{\Gamma}
\frac{m+\Gamma^0\mathcal{E}-\frac{b}{\sqrt{\,2}}\left(
\Gamma^1 Q+\Gamma^2 P\right)}
{m^2+\frac{b^2}{2}\left(Q^2+P^2-i\Gamma^1\Gamma^2\right)-\mathcal{E}^2}
\boldsymbol{\epsilon}(\mathbf{k}_1)\cdot\boldsymbol{\Gamma}, \\
\mathcal{E}&=&E_{N_i}^{(+)}-|\mathbf{k}_1|=E_{N_f}^{(+)}+|\mathbf{k}_2|.
\end{eqnarray*}
The operator $W(\mathcal{E})$ can be expressed in terms of the lowering and raising operators, 
\begin{eqnarray}
\label{W}
W(\mathcal{E})=-\Gamma^0
\left(\varepsilon_2^-\alpha+\varepsilon_2^+\alpha^\dagger\right)
\frac{m+\Gamma^0\mathcal{E}-i{b}\left(
\alpha a^\dagger+\alpha^\dagger a\right)}
{m^2+{b^2}\left(a^\dagger a+\alpha^\dagger \alpha\right)-\mathcal{E}^2}
\left(\varepsilon_1^-\alpha+\varepsilon_1^+\alpha^\dagger\right),
\end{eqnarray}
where
\[
\varepsilon_i^{\pm}=\varepsilon_i^1(\mathbf{k}_i)\pm 
i\varepsilon_i^2(\mathbf{k}_i), \qquad i=1,2.
\]

\paragraph{Factorization of the operators $e^{-i\mathbf{k}_i\cdot\mathbf{X}}$.} With the help of eqs.~(\ref{QP}) and (\ref{Q2P2}) the operator $e^{-i\mathbf{k}\cdot\mathbf{X}}$ can be represented as a product
\[
e^{-i\mathbf{k}_i\cdot\mathbf{X}}=V(\mathbf{k})\tilde V(\mathbf{k}), 
\]
where 
\begin{eqnarray}
\label{V}
V(\mathbf{k})&=&\exp\left[\frac 1 b \left(k^-a-k^+a^\dagger\right)\right], \qquad 
k^{\pm}=k^1\pm ik^2,\\
\tilde V(\mathbf{k})&=&\exp\left[-i\frac {\sqrt{\,2}} {b} \left(k^1\tilde Q-k^2\tilde P\right)\right].
\end{eqnarray}
The label $N$ of the state vector $\mid E>0, N\rangle$ is composite and consists of the label $n$ of the state vector in the space $\mathcal{H}(Q, P)\otimes\mathcal{H}(\Gamma)$ (see Appendix~A) and $\tilde n$ labelling the basis vectors in the space $\mathcal{H}(\tilde Q, \tilde P)$. 
We do not specify the latter basis here. The operators $\tilde V(\mathbf{k}_i)$ commute with $W(\mathcal{E})$. Moreover, the amplitude $T^{(2,1)}_{fi}$ factorizes:
\[
T^{(2,1)}_{fi}= 
\mathcal{T}^{(2,1)}_{fi}\;\tilde{\mathcal{T}}^{(2,1)}_{fi}, 
\] 
where
\[
\mathcal{T}^{(2,1)}_{fi}=\langle  E>0,\, n_f\mid
V(\mathbf{k}_2) W(\mathcal{E}) 
V(\mathbf{k}_1)\mid  E>0,\, n_i\rangle, \qquad
\tilde{\mathcal{T}}^{(2,1)}_{fi}=
\langle\tilde n_f \mid\tilde V(\mathbf{k}_2)\tilde V(\mathbf{k}_1)
\mid \tilde n_i\rangle.
\]
The matrix element $\tilde{\mathcal{T}}^{(2,1)}_{fi}$ can be easily calculated in a suitable basis (for example, in the oscillator basis in $\mathcal{H}(\tilde Q, \tilde P)$). However, it is, in some sense, irrelevant, since in the expression for the rate one has to perform a summation over all possible values of $\tilde{n}_f$, and this summation can be done without using the explicit form of $\tilde{\mathcal{T}}^{(2,1)}_{fi}$
and
$\tilde{\mathcal{T}}^{(1,2)}_{fi}$:
\begin{eqnarray*}
\sum_{\tilde{n}_f} |\mathcal{T}^{(2,1)}_{fi}\tilde{\mathcal{T}}^{(2,1)}_{fi}
+\mathcal{T}^{(1,2)}_{fi}\tilde{\mathcal{T}}^{(1,2)}_{fi}|^2
&=&|\mathcal{T}^{(2,1)}_{fi}|^2
+\mathcal{T}^{(2,1)}_{fi}{\mathcal{T}}^{(1,2)*}_{fi}
\exp\left[\frac{2i}{b^2}\left(\mathbf{k_1}\wedge
\mathbf{k}_2\right)\right]\\
&+&\mathcal{T}^{(1,2)}_{fi}{\mathcal{T}}^{(2,1)*}_{fi}
\exp\left[-\frac{2i}{b^2}\left(\mathbf{k_1}\wedge
\mathbf{k}_2\right)\right]
+|\mathcal{T}^{(1,2)}_{fi}|^2.
\end{eqnarray*}

Substituting the decomposition (\ref{positive}) of the state vectors
$\mid  E>0,\, n_i\rangle$, one finds  
\begin{eqnarray}
\label{calT}
&&\mathcal{T}^{(2,1)}_{fi}=
\frac{1}{2\sqrt{E_{n_f}E_{n_i}}}\nonumber\\
&&\times\Big\{\sqrt{(E_{n_f}+sm)(E_{n_i}+sm)}
\,\langle  n_f+1, s\mid
V(\mathbf{k}_2) W(\mathcal{E}) 
V(\mathbf{k}_1)\mid n_i+1, s\rangle\nonumber \\
&&+b\sqrt{\frac{E_{n_f}+sm}{E_{n_i}+sm}(n_i+1)}
\,\langle  n_f+1, s\mid
V(\mathbf{k}_2) W(\mathcal{E}) 
V(\mathbf{k}_1)\mid n_i, -s\rangle\nonumber \\
&&+b\sqrt{\frac{E_{n_i}+sm}{E_{n_f}+sm}(n_f+1)}
\,\langle  n_f, -s\mid
V(\mathbf{k}_2) W(\mathcal{E}) 
V(\mathbf{k}_1)\mid n_i+1, s\rangle \\.
&&+b^2\sqrt{\frac{(n_f+1)(n_i+1)}{(E_{n_f}+sm)(E_{n_f}+sm)}}
\,\langle  n_f, -s\mid
V(\mathbf{k}_2) W(\mathcal{E}) 
V(\mathbf{k}_1)\mid n_i, -s\rangle\Big\} \nonumber.
\end{eqnarray}

\paragraph{The matrix elements in $\mathcal{H}(\Gamma)$.} One can compute the matrix elements of the operator $W(\mathcal{E})$ in the basis $\mid \pm s\rangle$ in $\mathcal{H}(\Gamma)$. These matrix elements are operators in $\mathcal{H}(Q, P)$. Using the representation (\ref{W}) and eqs.~(\ref{gamma-s}), (\ref{alpha-s}), and (\ref{gamma-dagger}), one obtains
\begin{eqnarray}
&&\langle s\mid W(\mathcal{E})\mid s\rangle =\varepsilon_1^+\varepsilon_2^-
\left(\mathcal{E}-sm\right)\left[m^2+b^2(a^\dagger a+1)-\mathcal{E}^2\right]^{-1},\\
&&\langle s\mid W(\mathcal{E})\mid -s\rangle
=isb\varepsilon_1^-\varepsilon_2^-
\left[m^2+b^2(a^\dagger a+1)-\mathcal{E}^2\right]^{-1}a,\\
&&\langle -s\mid W(\mathcal{E})\mid s\rangle
=-isb\varepsilon_1^+\varepsilon_2^+ a^\dagger
\left[m^2+b^2(a^\dagger a+1)-\mathcal{E}^2\right]^{-1},\\
&&\langle -s\mid W(\mathcal{E})\mid -s\rangle
=\varepsilon_1^-\varepsilon_2^+
\left(\mathcal{E}+sm\right)\left[m^2+b^2a^\dagger a-\mathcal{E}^2\right]^{-1}.
\end{eqnarray}

\paragraph{The proper time integral.} Using the Schwinger proper-time representation,
\[
\left[m^2+b^2(a^\dagger a+1)-\mathcal{E}^2\right]^{-1}=i\int_0^\infty
d\lambda \exp\left\{-i\lambda\left[m^2+b^2(a^\dagger a+1)-\mathcal{E}^2\right]\right\} 
\]
and taking into account that 
\[
\mid n, \pm s\rangle= \mid n\rangle\otimes \mid \pm s\rangle
\]
one obtains
\begin{eqnarray}
\label{n+n+}
&&\langle n_2,\, s\mid V(\mathbf{k}_2) W(\mathcal{E})V(\mathbf{k}_1)\mid n_1,\, s\rangle \nonumber\\
&&=i\varepsilon_1^+\varepsilon_2^-
\left(\mathcal{E}-sm\right)
\int_0^\infty d\lambda 
\exp\left[-i\lambda\left(m^2+b^2-\mathcal{E}^2\right)\right]
\\
&&\phantom{=}\times\langle n_2\mid V(\mathbf{k}_2)\exp\left(-i\lambda b^2a^\dagger a\right) V(\mathbf{k}_1)\mid n_1\rangle,\nonumber
\\ 
\nonumber \\
\label{n+n-}
&&\langle n_2,\, s\mid V(\mathbf{k}_2) W(\mathcal{E})
V(\mathbf{k}_1)\mid n_1,\, -s\rangle \nonumber \\ &&=-sb\varepsilon_1^-\varepsilon_2^-
\int_0^\infty d\lambda
\exp\left[-i\lambda\left(m^2+b^2-\mathcal{E}^2\right)\right]\\
&&\phantom{=}\times\langle n_2\mid V(\mathbf{k}_2)\exp\left(-i\lambda b^2a^\dagger a\right)a V(\mathbf{k}_1)\mid n_1\rangle, \nonumber
\\
\nonumber \\
\label{n-n+}
&&\langle n_2, \, -s\mid V(\mathbf{k}_2) W(\mathcal{E})
V(\mathbf{k}_1)\mid n_1, \, s\rangle \nonumber\\
&&=sb\varepsilon_1^+\varepsilon_2^+
\int_0^\infty d\lambda
\exp\left[-i\lambda\left(m^2+b^2-\mathcal{E}^2\right)\right]        \\
&&\phantom{=}\times\langle n_2\mid V(\mathbf{k}_2)\,a^\dagger\exp\left(-i\lambda b^2a^\dagger a\right) V(\mathbf{k}_1)\mid n_1\rangle,\nonumber
\\
\nonumber \\
\label{n-n-}
&&\langle n_2, \,-s\mid V(\mathbf{k}_2)
W(\mathcal{E})V(\mathbf{k}_1)\mid n_1,\,-s\rangle \nonumber \\
&&=i\varepsilon_1^-\varepsilon_2^+\left(\mathcal{E}+sm\right)
\int_0^\infty d\lambda
\exp\left[-i\lambda\left(m^2-\mathcal{E}^2\right)\right] \\
&&\phantom{=}\times\langle n_2\mid V(\mathbf{k}_2)\exp\left(-i\lambda b^2a^\dagger a\right) V(\mathbf{k}_1)\mid n_1\rangle. \nonumber
\end{eqnarray}
Using eqs.~(\ref{V}) and (\ref{bosonic}) one easily derives
\begin{equation}
\label{aV}
[a, V(\mathbf{k})]_-=-\frac{k^+}{b}V(\mathbf{k}), \qquad
[a^\dagger, V(\mathbf{k})]_-=-\frac{k^-}{b}V(\mathbf{k}).
\end{equation}
Substituting eqs.~(\ref{n+n+})-(\ref{n-n-}) in eqs.~(\ref{calT}) and using eqs.~(\ref{aV}), one can put the amplitude $\mathcal{T}_{fi}^{(2,1)}$ into the form
\begin{eqnarray}
\label{calT1}
&&\mathcal{T}^{(2,1)}_{fi}=
\frac{1}{2\sqrt{E_{n_f}E_{n_i}}}\int_{0}^\infty d\lambda
\exp\left[-i\lambda \left(
m^2+b^2-\mathcal{E}^2\right)\right]
\nonumber\\
&&\times\Bigg\{i\varepsilon_1^+\varepsilon_2^-(\mathcal{E}-sm)
\sqrt{(E_{n_f}+sm)(E_{n_i}+sm)}
\,\langle  n_f+1\mid
D(\mathbf{k}_2, \mathbf{k}_1)\mid n_i+1\rangle
\nonumber \\
&&-sb\varepsilon_1^-\varepsilon_2^-\sqrt{\frac{E_{n_f}+sm}{E_{n_i}+sm}(n_i+1)}
\Big[\,b\langle  n_f+1\mid
D(\mathbf{k}_2, \mathbf{k}_1)\mid n_i-1\rangle
\nonumber\\
&&\phantom{xxxxxxxxxxxxxxxxxxxxxx}
-k_1^+\langle  n_f+1\mid
D(\mathbf{k}_2, \mathbf{k}_1)\mid n_i\rangle\Big]
\\
&&+sb\varepsilon_1^+\varepsilon_2^+\sqrt{\frac{E_{n_i}+sm}{E_{n_f}+sm}(n_f+1)}
\Big[\,b\langle  n_f-1\mid
D(\mathbf{k}_2, \mathbf{k}_1)\mid n_i+1\rangle
\nonumber \\
&&\phantom{xxxxxxxxxxxxxxxxxxxxxx}
+k_2^-\langle  n_f\mid
D(\mathbf{k}_2, \mathbf{k}_1)\mid n_i+1\rangle\Big]
\nonumber \\
&&+ib^2e^{i\lambda b^2}\varepsilon_1^-\varepsilon_2^+(\mathcal{E}-sm)
\sqrt{\frac{(n_f+1)(n_i+1)}{(E_{n_f}+sm)(E_{n_f}+sm)}}
\,\langle  n_f\mid
D(\mathbf{k}_2,\mathbf{k}_1)\mid n_i\rangle\Bigg\}, \nonumber
\end{eqnarray}
where $D(\mathbf{k}_2,\mathbf{k}_1)=V(\mathbf{k}_2)\exp(-i\lambda b^2a^\dagger a)
V(\mathbf{k}_1)$.  The  
matrix element $\langle  n_f-1\mid
D(\mathbf{k}_2, \mathbf{k}_1)\mid n_i+1\rangle$ in (\ref{calT1}) must be replaced by zero for transitions to the ground state.

\paragraph{The matrix elements in $\mathcal{H}(Q, P)$.}
We are going to compute the matrix elements in the space $\mathcal{H}(Q, P)$ in eq.~(\ref{calT1}). 
The operator $V(\mathbf{k})$ has simple form in the coherent state basis (this is widely used in quantum optics). A harmonic oscillator coherent state \cite{Glauber63}, see, e.g.,~\cite{ZhFG90}, is given by
\begin{equation}
\label{coherent}
\mid z\rangle=\exp(za^\dagger-z^*a)\mid 0\rangle
=\sum_{n=0}^{\infty}\frac{z^n}{\sqrt{n!}}\mid n\rangle 
\exp\left(-\frac 1 2 z^* z\right),
\end{equation}
where $\mid 0\rangle$ is the harmonic oscillator vacuum (\ref{vacuum}) and $z$ is a complex number. 
Using eqs.~(\ref{V}) and (\ref{coherent}), one derives
\begin{eqnarray}
&&V(\mathbf{k})\mid z \rangle=\mid z-\frac{k^+}{b}\rangle
\exp\left(\frac{k^-z-k^+z^*}{2b}\right)\\
&&\langle z \mid V(\mathbf{k})=\exp\left(\frac{k^-z-k^+z^*}{2b}\right)\langle z+\frac{k^+}{b}\mid.
\end{eqnarray}
Then, using the properties of the coherent states \cite{ZhFG90}, one finds that the matrix elements are represented as
\begin{eqnarray}
\label{mel}
&&\langle n_2\mid V(\mathbf{k}_2)\exp\left(-i\lambda b^2a^\dagger a\right) V(\mathbf{k}_1)\mid n_1\rangle\nonumber
\\
&&= \frac{1}{\sqrt{n_1!n_2!}}\exp\left\{-\frac{1}{2b^2}\left[
\mathbf{k}_1^2+\mathbf{k}_2^2+2(\mathbf{k}_1\cdot\mathbf{k}_2
-i\mathbf{k}_1\wedge\mathbf{k}_2)e^{-i\lambda b^2}
\right]\right\}\nonumber
\\
&&\phantom{=}\times J^{(0,n_2,n_1,0)}
\left(\frac{k_1^-+k_2^-e^{-i\lambda b^2}}{b}, 0, 0,
-\frac{k_2^++k_1^+e^{-i\lambda b^2}}{b} 
\right)
\end{eqnarray}
where
\begin{eqnarray*}
&&J^{(l_1,l_2,l_3,l_4)}(\zeta_1,\,\zeta_2,\,\zeta_3,\,\zeta_4)=
\prod_{i=1}^4\left(\frac{\partial}{\partial\zeta_i}\right)^{l_i}
J(\zeta_1,\,\zeta_2,\,\zeta_3,\,\zeta_4),
\end{eqnarray*}
the generating function $J(\zeta_1,\,\zeta_2,\,\zeta_3,\,\zeta_4)$ is given by
\begin{eqnarray*}
&&J(\zeta_1,\,\zeta_2,\,\zeta_3,\,\zeta_4)=\frac{1}
{(2\pi i)^2}\int dz_2^* dz_2 dz_1^* dz_1 \exp\big(-z_1^* z_1
-z_2^* z_2
\\
&&\phantom{xxxxxxxxxxxxxx}
+z_2^*z_1e^{-i\lambda b^2}+\zeta_1 z_1+
\zeta_2 z_2+\zeta_3 z_1^*+\zeta_4 z_2^*\big).
\end{eqnarray*}
A direct calculation yields 
\begin{equation}
\label{J0}
J(\zeta_1,\,\zeta_2,\,\zeta_3,\,\zeta_4)=\exp\left(
\zeta_1\zeta_3+\zeta_2\zeta_4+\zeta_2\zeta_3e^{-i\lambda b^2}
\right).
\end{equation}
Introducing the dimensionless momenta 
\begin{equation}
\label{kappa}
\boldsymbol{\kappa}_i=b^{-1}\mathbf{k}_i, \qquad
i=1,2; \qquad 
\kappa_{i\pm}=b^{-1}k_i^\pm,
\end{equation}
and using eq.~(\ref{J0}), one obtains, for $n_2\le n_1$,   
\begin{eqnarray}
\label{J0nn0}
&&J^{(0,n_2,n_1,0)}
\left(\kappa_{1-}+\kappa_{2-}e^{-i\lambda b^2}, 0, 0,
-\kappa_{2+}-\kappa_{1+}e^{-i\lambda b^2}
\right)
\nonumber \\
&&=\sum_{l=0}^{n_2}\sum_{r=0}^{n_1-n_2+l}\sum_{t=0}^{l}
\frac{(-)^{l}\,n_1!n_2!}{(n_2-l)!\,r!\,(n_1-n_2+l-r)!\,t!\,(l-t)!}
\nonumber \\
&&\phantom{=}\times \kappa_{1+}^t\kappa_{1-}^{r}
\kappa_{2+}^{l-t}\kappa_{2-}^{n_1-n_2+l-r} 
\exp\left[-i\lambda b^2(n_1-r+t)\right].
\end{eqnarray}
Substituting eqs.~(\ref{J0nn0}) and (\ref{kappa}) in eq.~(\ref{mel}), one finds
\begin{eqnarray}
\label{mel1}
&&\langle n_2\mid D(b\boldsymbol{\kappa}_2,b\boldsymbol{\kappa}_1)\mid n_1\rangle=
\langle n_2\mid V(b\boldsymbol{\kappa}_2)\exp\left(-i\lambda b^2a^\dagger a\right) V(b\boldsymbol{\kappa}_1)\mid n_1\rangle\nonumber
\\
&&= \sqrt{n_1!n_2!}\,\exp\left\{-\frac{1}{2}\left(
\boldsymbol{\kappa}_1^2+\boldsymbol{\kappa}_2^2\right)-\left[\boldsymbol{\kappa}_1\cdot\boldsymbol{\kappa}_2
-i\boldsymbol{\kappa}_1\wedge\boldsymbol{\kappa}_2 \exp(-i\lambda b^2)
\right]\right\}\nonumber\\
&&\times\sum_{l=0}^{n_2}\sum_{r=0}^{n_1-n_2+l}\sum_{t=0}^{l}
\frac{(-)^{l}}{(n_2-l)!\,r!\,(n_1-n_2+l-r)!\,t!\,(l-t)!}
\nonumber \\
&&\phantom{=}\times \kappa_{1+}^t\kappa_{1-}^{r}
\kappa_{2+}^{l-t}\kappa_{2-}^{n_1-n_2+l-r} 
\exp\left[-i\lambda b^2(n_1-r+t)\right].
\end{eqnarray}
One has to substitute this expression in eq.~(\ref{calT1}). The general formula is large and we do not present it here.  
The transition amplitude from the second to the first (fundamental) Landau level can serve as an example.
The amplitude $\mathcal{T}^{(2,1)}_{01}$ is given by  
\begin{eqnarray}
\label{10}
&&\mathcal{T}^{(2,1)}_{01}=\frac{e^{-i\theta_1}}{2\sqrt{E_{0}E_{1}}}\exp\left(-\frac{{\kappa}_1^2+{\kappa}_2^2}{2}\right)\sum_{n=0}^\infty\frac{\left(-{\kappa}_1{\kappa}_2 e^{-i\phi}\right)^n}{n!} 
\nonumber\\
&&\times\Bigg\{(\mathcal{E}-sm)
\sqrt{\frac{(E_{0}+sm)(E_{1}+sm)}{2}}
\Big[-\frac{{\kappa}_1^2{\kappa}_2}{E_n^2-\mathcal{E}^2}
\nonumber\\
&&+\frac{{\kappa}_1\left(2-{\kappa}_1^2-2{\kappa}_2^2\right)}{E_{n+1}^2-\mathcal{E}^2}e^{-i\phi}
+\frac{{\kappa}_2\left(2-2{\kappa}_1^2-{\kappa}_2^2\right)}{E_{n+2}^2-\mathcal{E}^2}e^{-2i\phi}
-\frac{{\kappa}_1{\kappa}_2^2}{E_{n+3}^2-\mathcal{E}^2}e^{-3i\phi}
\Big]
\nonumber \\
&&+isb^2\sqrt{\frac{2\left(E_{0}+sm\right)}{E_{1}+sm}}
\Big[\frac{{\kappa}_2\left(1-{\kappa}_1^2\right)}{E_n^2-\mathcal{E}^2}
+\frac{{\kappa}_1\left(2-{\kappa}_1^2-{\kappa}_2^2\right)}{E_{n+1}^2-\mathcal{E}^2}e^{-i\phi}
-\frac{{\kappa}_1{\kappa}_2}{E_{n+2}^2-\mathcal{E}^2}e^{-2i\phi}\Big]
\nonumber\\
&&+isb^2\sqrt{\frac{E_{1}+sm}{2\left(E_{0}+sm\right)}}
\Big[\frac{{\kappa}_1^2{\kappa}_2}{E_{n}^2-\mathcal{E}^2}
+\frac{2{\kappa}_1{\kappa}_2^2}{E_{n+1}^2-\mathcal{E}^2}e^{-i\phi}
+\frac{{\kappa}_2^3}{E_{n+2}^2-\mathcal{E}^2}e^{-2i\phi}\Big]
\nonumber\\
&&+b^2\left(\mathcal{E}+sm\right)
\sqrt{\frac{2}{(E_{0}+sm)(E_{1}+sm)}}\Big[
\frac{{\kappa}_1}{E_{n-1}^2-\mathcal{E}^2}e^{i\phi}+
\frac{{\kappa}_2}{E_{n}^2-\mathcal{E}^2}\Big]\Bigg\}
\end{eqnarray}
where $\theta_i$ is the angle between the $x$-axes and $\boldsymbol{\kappa}_i$, $\phi=\theta_2-\theta_1$, $\kappa_i=|\boldsymbol{\kappa}_i|$, and we have put $E_{-1}=m$.
The second part of the amplitude is obtained by the replacements
\[
\kappa_1\leftrightarrow\kappa_2, \quad 
\theta_1\leftrightarrow\theta_2,\quad
\phi\rightarrow -\phi, \quad \mathcal{E}\rightarrow
\mathcal{E}^{\prime}=E_0+k_2.
\] 
The sum in eq.~(\ref{10}) can be taken and the result can be expressed in terms of the confluent hypergeometric function \cite{Herold79}.

\section{Conclusions}

Within the framework of the Schwinger proper-time method the dynamics of a fermion in a constant magnetic field in $2+1$ dimensional space-time reduces to that of a supersymmetric quantum mechanical system with one bosonic and one fermionic degrees of freedom. An expression is obtained for the two-photon emission amplitude. Similar technique can be used in $3+1$ dimensions as well. The work on this case is in progress.  


\appendix
\section{Eigenvalues and eigenstates of the Dirac \\hamiltonian in a constant magnetic field}

The Dirac hamiltonian in the abstract Hilbert space  reads
\begin{equation}
\label{hamQP}
{H}=\Gamma^0\left[\frac{b}{\sqrt{\,2}}\left(\Gamma^1 Q+ \Gamma^2 P\right)+m\right],
\end{equation}
where $Q$ and $P$ are defined in eq.~(\ref{QP}). It actually acts in the factor $\bar\mathcal{H}=\mathcal{H}(Q,P)\otimes\mathcal{H}(\Gamma)$. 
The hamiltonian (\ref{hamQP}) can be expressed in terms of the raising and lowering operators (\ref{creation}) and (\ref{creationF}), 
\begin{equation}
\label{hamalphaa}
{H}=\Gamma^0\left[ib\left(\alpha a^\dagger+\alpha^\dagger a\right)+m\right].
\end{equation}
The operators $a$, $a^\dagger$ act in the factor $\mathcal{H}(Q,P)$, while $\alpha$, $\alpha^\dagger$ act in the factor $\mathcal{H}(\Gamma)$.

The operator $\Gamma^0$ has eigenvalues $\pm s$ and both eigenspaces in $\mathcal{H}(\Gamma)$ are one-dimensional. Let $\mid s\rangle$ be a normalized eigenvector of $\Gamma^0$ in $\mathcal{H}(\Gamma)$ corresponding to the eigenvalue $s$, 
\begin{equation}
\label{gamma-s}
\Gamma^0\mid s\rangle=s\mid s\rangle, 
\qquad \langle s\mid s\rangle=1.
\end{equation}
It is easy to check, using eqs.~(\ref{creationF}), that
\begin{equation}
\label{alpha-s}
\alpha\mid s\rangle=0, 
\end{equation}
and the vector $\mid -s\rangle=\alpha^\dagger\mid s\rangle$ satisfies
\begin{equation}
\label{gamma-dagger}
\Gamma^0\mid -s\rangle=-s\mid -s\rangle, \qquad
\langle -s\mid -s\rangle=1, \qquad
\alpha^\dagger\mid -s\rangle=0.
\end{equation}
The vectors $|\pm s\rangle$ form an orthonormal basis in $\mathcal{H}(\Gamma)$. 
On the other hand, the vectors 
\[
\mid n \rangle= \frac{(a^\dagger)^n}{\sqrt{n!}}
\mid 0\rangle, \qquad n=0,1,\dots,
\]
where 
\begin{equation}
\label{vacuum}
a\mid 0 \rangle=0, \qquad \langle 0\mid 0\rangle=1,
\end{equation}
form an orthonormal basis in $\mathcal{H}(Q,P)$, as is well known from the harmonic oscillator theory. Then an orthonormal basis in $\bar\mathcal{H}$ is given by the set of the vectors
\[
\mid n, \pm s\rangle=\mid n \rangle\,\otimes\mid \pm s \rangle, 
\qquad n=0, 1, \dots,
\]
for which the following relations hold: 
\begin{eqnarray}
\label{a}
&&a\mid 0, \pm s\rangle=0, \qquad
a\mid n, \pm s\rangle=\sqrt{n}\mid n-1, \pm s\rangle, \qquad
n=1,2,  \dots;  \\
&&a^\dagger\mid n, \pm s\rangle=\sqrt{n+1}\mid n+1, \pm s\rangle,  \\
&&\alpha\mid n, s\rangle=0, \qquad
\alpha^\dagger\mid n, s\rangle=\mid n, -s\rangle, \\
&&\alpha^\dagger\mid n, -s\rangle=0, \qquad
\alpha\mid n, -s\rangle=\mid n, s\rangle, \hspace{26mm} n=0,1, \dots.
\end{eqnarray}

Let $\mid\psi\rangle$ be an eigenvector of $H$ with eigenvalue $E$, 
\begin{equation}
\label{eveq}
H\mid\psi\rangle=E\mid\psi\rangle.
\end{equation}
Applying  $H$ to both sides of this equation and using eqs.~(\ref{hamQP}), (\ref{creation}), and (\ref{creationF}), one obtains
\[
\left[b^2\left(a^\dagger a+\alpha^\dagger\alpha\right)
+m^2\right]\mid \psi\rangle=E^2\mid\psi\rangle.
\] 
Using the decomposition
\[
\mid\psi\rangle=\mid\psi_s\rangle+\mid\psi_{-s}\rangle,
\]
where $\alpha\mid\psi_s\rangle=0$,  
$\alpha^\dagger\mid\psi_{-s}\rangle=0$
and eq.~(\ref{anti}), one finds 
\begin{eqnarray}
\label{psi1}
&&b^2a^\dagger a\mid\psi_s\rangle=(E^2-m^2)\mid\psi_s\rangle, \\
\label{psi2}
&&b^2a^\dagger a\mid\psi_{-s}\rangle=(E^2-m^2-b^2)\mid\psi_{-s}\rangle.
\end{eqnarray}
So, both components $\psi_{\pm s}$ must be eigenvectors of $a^\dagger a$.
Using (\ref{psi1}), (\ref{psi2}) one finds the squared eigenvalues, 
\begin{equation}
\label{value}
E_n^2=m^2+(n+1)b^2, \qquad n=0, 1,\dots
\end{equation}
and the form of the eigenvectors
\begin{equation}
\label{evec}
\mid\psi\rangle=c_s\mid n+1, s\rangle+c_{-s}\mid n, -s\rangle.
\end{equation}
Inserting (\ref{evec}) into (\ref{psi1}), (\ref{psi2}) and using (\ref{value}) one obtains for the eigenvectors, after some simple algebra,  
\begin{eqnarray}
\label{positive}
\mid E>0, n\rangle= \frac 1 {\sqrt{2E_n}}\left[\sqrt{E_n+sm}\mid n+1, s\rangle+b\sqrt{\frac{n+1}{E_n+sm}}\mid n,-s\rangle\right], \\
\label{negative}
\mid E<0, n\rangle= \frac 1 {\sqrt{2E_n}}\left[\sqrt{E_n-sm}\mid n+1, s\rangle-b\sqrt{\frac{n+1}{E_n-sm}}\mid n,-s\rangle\right],\\
n=0,1,2\dots . \nonumber
\end{eqnarray}
The eigenvalues are $\pm E_n$, correspondingly, where $E_n=\sqrt{E_n^2}$.

\begin{acknowledgments}
We are indebted to V.~G.~Bagrov for discussions and useful remarks.

The authors thank the referees for many valuable comments.

The authors thank Conselho Nacional de Desenvolvimento Cient\'{i}fico e Tecnol\'{o}gico (CNPq) of Brazil for financial support.

\end{acknowledgments}
\bibliography{phem5-5}
\end{document}